\documentstyle[12pt]{article}
\newcommand{\idty}{{\leavevmode{\rm 1\mkern -5.4mu I}}}
\newcommand{\Ibb}[1]{ {\rm I\ifmmode\mkern
            -3.6mu\else\kern -.2em\fi#1}}
\newcommand{\ibb}[1]{\leavevmode\hbox{\kern.3em\vrule
     height 1.2ex depth -.3ex width .2pt\kern-.3em\rm#1}}

\begin{document}

\centerline{\Large \bf A note on  the lattice
Dirac-K\"ahler equation}
\vskip1.cm
\begin{center}
\begin{minipage}{13cm}
 {\bf Timothy Striker}\\
Institut f\"ur Theoretische Physik, Bunsenstr. 9, D-37073
G\"ottingen\\
email: striker@theorie.physik.uni-goettingen.de
\end{minipage}
\end{center}
\vskip1.cm

\begin{abstract}
\noindent
A lattice version of the Dirac-K\"ahler equation (DKE) describing
fermions was discussed in articles by Becher and Joos. The
decomposition of lattice Dirac-K\"ahler fields (inhomogeneous cochains)
to lattice Dirac fields remained as an open problem.
I show that it is possible to extract Dirac fields from the DKE and
discuss the resulting lattice Dirac equation.
\end{abstract}

\section{Introduction}

In 1962 E. K\"ahler proposed an alternative to  the Dirac equation
describing fermionic fields \cite{Kahler}.
In his framework the fermionic fields are inhomogeneous differential forms
\begin{equation}
 {\bf \phi}= \phi^0(x)+ \phi^1_{i}(x) dx^i +
              \phi^2_{ij}(x) dx^i dx^j +
              \phi^3_{ijk}(x) dx^i dx^j dx^k  +
              \phi^4(x) dx^1 dx^2 dx^3 dx^4
\end{equation}
obeying the  Dirac-K\"ahler equation (DKE)
\begin{equation}
  \label{DKE}
  (d-\delta+m){\bf \phi}=0\, .
\end{equation}
Here $d$ is the ordinary exterior derivative and $\delta$ is the
coderivative $\delta= - \ast^{-1} d\, \ast$\, (where $\ast$ is the
Hodge-operator).

Similar to the ordinary Dirac-operator acting on spinors the operator
$d-\delta$ is a root of the Laplace (or Laplace-Beltrami) operator
acting on inhomogeneous differential forms.
\begin{equation}
  (d-\delta)^2 = - d \delta - \delta d = \partial_i\partial^i\, .
\end{equation}
Here and in the following I use the summation convention on space-time
indices. I work in a four-dimensional Euclidean space-time, but
changing the signature to Minkowski and going to other space-time
dimensions is unproblematic.

By introducing a suitable basis of the space of inhomogeneous
differential forms the DKE can be seen to be equivalent to four
degenerate Dirac equations \cite{BecherJoos}.

Becher and Joos use a correspondence between the continuum and the
lattice which has its origin in algebraic topology (see for example
Nakahara \cite{Nakahara}) to discretize the DKE on a hypercubic lattice.
In this correspondence differential forms are related to cochains
(functions on the points, links, plaquettes, cubes and
hypercubes of the lattice). Thus the lattice Dirac-K\"ahler field is an
inhomogeneous cochain. The lattice DKE reads
\begin{equation}
  (\tilde\Delta-\tilde\nabla+m) {\bf\phi}=0
\end{equation}
where $\tilde\Delta$ denotes the dual boundary and $\tilde\nabla$ the
dual coboundary operator. The lattice DKE shares the property of the
continuum DKE of being a root of the (lattice) Klein-Gordon equation.
The procedure to read off Dirac components and a Dirac
equation from the DKE does not extend directly to the lattice. Finding a
lattice version of this procedure remained an open problem
(see the corresponding remark in \cite{Joos}). The reduction presented
in \cite{BecherJoos} works only in momentum space.

The inhomogeneous cochains $\bf \phi$ can be expanded in the cochains
 $d^{x,H}$
\begin{equation}
  {\bf \phi}=\sum_{x,H} \varphi(x,H)\, d^{x,H}\, ,
\end{equation}
where the $H$ are  sets $\{i_1\dots i_p\}$ \, with
$p=1,\dots, 4$ and $x$ is a point of the lattice.
The index sets $(x,H)$ label the points, links, plaquettes, cubes and
hypercubes of the lattice.
The $d^{x,H}$ are antisymmetric in their indices
$d^{x,i_1\dots i_{k+1}i_k\dots i_p}= -\, d^{x,i_1\dots i_ki_{k+1}\dots i_p}$.
The subset of the $ d^{x,H}$ where $H$ is an ordered set is a basis
of the space of cochains.

We can  sum up the $d^{x,H}$  over the lattice to get constant cochains $d^H$.
\begin{equation}
  d^H=\sum_x d^{x,H}
\end{equation}
{}From these the original $d^{x,H}$ can be recovered by
\begin{equation}
  d^{x,H} = \chi_{(x,H)} d^H
\end{equation}
where $\chi$ is the characteristic function for a lattice with spacing $a$
\begin{equation}
  \chi_{(x,H)}(y) = \left\{
  \begin{array}{l}
    1/a^p \qquad\mbox{if}\quad y\in (x,H)\\ 0 \qquad
    \mbox{otherwise}\, .
  \end{array}\right.
\end{equation}
A general cochain $\phi$ can then be written as
\begin{equation}
  \begin{array}{rl}
    \phi= \displaystyle\sum_{x,H} \varphi(x,H)\, d^{x,H} \!\!
    &=\displaystyle\sum_{x,H} \varphi(x,H) \chi_{(x,H)}\, d^H
    \qquad\\[1.9ex] &=\displaystyle\sum_H \left(\sum_x \varphi(x,H)
    \chi_{(x,H)}\right)\, d^H = \displaystyle\sum_H \varphi_H(x)\,d^H\, .
  \end{array}
\end{equation}

The dual boundary and coboundary operator act on a p-cochain
\begin{equation}
  {\bf \omega} = \frac{1}{p!}\sum_{i_1\dots i_p}
   \omega_{i_1\dots i_p}(x)  \, d^{i_1\dots i_p}
\end{equation}
via
\begin{equation}
  \tilde\Delta  {\bf \omega} = \frac{1}{p!}\sum_{j i_1\dots i_p}
   \partial_{+j}\omega_{i_1\dots i_p}(x)  \,d^{j i_1\dots i_p}
\quad , \quad
 \tilde\nabla  {\bf \omega} = \frac{1}{(p-1)!}\sum_{j i_2\dots i_p}
   \partial_{-j} \omega_{ji_2\dots i_p}(x)  \, d^{i_2\dots i_p}
\end{equation}
where $\partial_{+i}$ and $\partial_{-i}$ are the lattice (nearest
neighbor) derivatives
\begin{equation}
 \partial_{+i}f(x)=\frac{1}{a}(f(x+a^i)-f(x))\qquad
 \partial_{-i}f(x)=\frac{1}{a}(f(x)-f(x-a^i))
\end{equation}
with $(x+a^i)^j= x^j+ \delta^{ij} a$.

\section{Dirac-components for the lattice DKE}

To read off Dirac-components from the lattice DKE I have to introduce
an algebra of matrices $C^i$, $i=1,\dots,4$, replacing the
Dirac-$\gamma$-matrices on the lattice.
The $C^i$ are  a representation of the anti-commutation relations
\begin{equation}\label{cmatrices}
\{ C^i\, , \, C^j \,\}=0 \qquad \{ C^i\, ,\, C^j{}^\dagger \,\}=\delta^{ij}\,
.\end{equation}
The lattice Dirac equation involving the $C^i$ reads
\begin{equation}\label{dirac}
\bigl( C^i \partial_{+i} + C^i{}^\dagger\partial_{-i} + m \idty\bigr)\psi
=0\, .
\end{equation}
The usual lattice Klein-Gordon equation \cite{MontvayMunster} is
obtained by multiplying (\ref{dirac}) with
$ C^i \partial_{+i} + C^i{}^\dagger\partial_{-i} - m$ just as in the
continuum \footnote{This is not the case for the discretizations of the
Dirac equation in which the $\gamma^i$ appear.}. An equivalent version
of the lattice Dirac equation was first discussed by Becher
\cite{Becher}. He shows that this equation does not lead to the
well known fermion doubling problem  from additional zeros in the
inverse propagator \cite{MontvayMunster}.

A drawback of this treatment of fermions on the lattice appears if
we try to find a representation of the algebra (\ref{cmatrices}).  The
eight combinations
\begin{equation}
  \xi^i=  C^i+ C^i{}^\dagger
      \quad,\qquad
\xi^{4+i}= i\, \left( C^i{}^\dagger- C^i\right)
\end{equation}
fulfill the defining relations of an 8 generator Clifford algebra
\begin{equation}\label{Clifford}
  \{ \xi^i, \xi^j\}= 2 \delta^{ij}\, .
\end{equation}
Therefore the smallest representation of the $C^i$ is 16-dimensional
\cite{Cornwell}. For any representation $\gamma^i$ of the Dirac
algebra the matrices $\xi^i= \idty \otimes \gamma^i$,
$\xi^{4+i}= \gamma^i \otimes\gamma^5$ are a representation of
(\ref{Clifford}). In this representation it is particularly easy to
see what happens in the continuum limit. For the lattice derivatives
we have $\partial_{+i}\longrightarrow \partial_i$ and
$\partial_{-i}\longrightarrow \partial_i$. For the Dirac operator
this leads to
\begin{equation}
C^i \partial_{+i} + C^i{}^\dagger\partial_{-i} + m \,
-\!\!\!-\!\!\!\longrightarrow  \,\,\xi^i \partial_i + m   \, .
\end{equation}
Obviously our lattice Dirac equation is equivalent to four independent
Dirac equations in the continuum limit. Thus the fermion doubling that was
absent in the spectrum  partially comes back with the size of the
representation. This seems strange if we are interested in discretizing
the ordinary Dirac equation, but it is natural if we think of the
Dirac-K\"ahler equation, because as mentioned above this equation
describes four independent fermions in the first place.

To read off Dirac components from the lattice DKE a suitable basis of the
space of inhomogeneous cochains has to be introduced. To do so I
notice that the algebra (\ref{cmatrices}) is the algebra of
fermionic creation and annihilation operators. Therefore  a
vector $\Omega$ with $C^i{}^\dagger\,\Omega=0$ for all $C^i{}^\dagger$
exists. With the help of this let us define
\begin{equation}
\theta = \Omega^\dagger \left(\idty + d^i C^i{}^\dagger +\frac{1}{2!}\, d^{ij}
  C^j{}^\dagger C^i{}^\dagger
+\frac{1}{3!}\, d^{ijk}  C^k{}^\dagger C^j{}^\dagger C^i{}^\dagger
+\frac{1}{4!}\, d^{ijkl}
  C^l{}^\dagger C^k{}^\dagger C^j{}^\dagger C^i{}^\dagger\right)\, .
\end{equation}
This equation is to be
read as a 16-dimensional vector equation. To prove that the components
of $\theta$ are a basis of the space of inhomogeneous differential
forms I give the inverse of this formula,
\begin{equation}
   d^H = \theta\,C^H \Omega
\end{equation}
with $C^H= C^{i_1}\cdots  C^{i_p}$ for $H=\{i_1\dots i_p\}$.
(Here I assume that $\Omega$ is normalized $\Omega^\dagger\Omega=1$.)
A solution of the lattice DKE can be expanded in terms of this basis
(in vector notation)
\begin{equation}
  \phi = \theta \cdot \varphi(x)
\end{equation}

A somewhat lengthy calculation reveals that  DKE in the basis $\theta$ reads
\begin{equation}
  (d-\delta+m) \, \theta\cdot \varphi(x) = \theta\cdot \bigl(C^i
  \partial_i + C^i{}^\dagger \partial_{-i}+m \bigr)\varphi(x)\, .
\end{equation}
So we see that the components  $\varphi$ are Dirac components for the
lattice DKE. They obey Becher's version of the lattice Dirac equation.
In the continuum limit we get four independent Dirac equations.

\section{Summary}
\setcounter{equation}{0}

I have demonstrated that it is possible to read off Dirac components from
the lattice DKE. In the continuum case the equation for the 16 Dirac
components of the Dirac-K\"ahler field can be reduced to 4 independent
4-component Dirac equations.  This is not possible on the lattice. For
nonzero lattice spacing we get one irreducible 16 component
Dirac equation. Only in the continuum limit the continuum result is
reproduced and we get four independent equations.

The resulting lattice Dirac equation is discussed by Becher in his
paper \cite{Becher}. Here I only want to point out that there is no
further fermion-doubling from the spectrum of the discrete Dirac
operator and that there is a discrete $\gamma^5$ invariance (for the
question of chiral symmetry see also \cite{Rabin}).

\vspace*{4ex}

Thanks are due to  Folkert M\"uller-Hoissen and Aristophanes Dimakis
for numerous discussions.

\end{document}